\documentclass[conference,10pt,letterpaper]{IEEEtran}

\usepackage{cite}
\usepackage{epsfig}
\usepackage{epstopdf}
\usepackage{graphicx}
\usepackage[dvipsnames]{xcolor}
\usepackage{tikz}
\usetikzlibrary{arrows,positioning,shapes.geometric,circuits.logic.US}
\usepackage{pgfplots}
\usepackage{multirow}
\usepackage{upgreek}
\usepackage{amssymb}
\usepackage{amsmath}
\usepackage{cases}
\usepackage{url}
\usepackage{pifont}
\usepackage{bm}
\usepackage{amsthm}

\usepackage{tabularx}
\usepackage{booktabs}
\usepackage{filecontents}
\usepackage{subcaption}
\newcolumntype{Y}{>{\centering\arraybackslash}X}

\theoremstyle{definition} 
\theoremstyle{definition} 
\theoremstyle{definition} 
\theoremstyle{definition}

\newcolumntype{M}[1]{>{\centering\arraybackslash}m{#1}}

\usepackage[linesnumbered, ruled, vlined, noend]{algorithm2e}

\usepackage{algorithmic}

\usepackage{array}

\newcommand{\fixme}[2]{\ifx&#2&{\leavevmode\color{red}#1}\else{\leavevmode\color{red}FIXME\{}#1{\leavevmode\color{red}\}}\footnote{{\leavevmode\color{red}#2}}\PackageWarning{Fixme}{#1: #2}\fi}

\newcommand{\newstuff}[2]{\ifx&#2&{\leavevmode\color{blue}#1}\else{\leavevmode\color{blue}FIXME\{}#1{\leavevmode\color{blue}\}}\footnote{{\leavevmode\color{blue}#2}}\PackageWarning{Newstuff}{#1: #2}\fi}

\title{Iterative Soft-Input Soft-Output Decoding with Ordered Reliability Bits GRAND}

\author{\IEEEauthorblockN{Carlo~Condo}}

\begin{document}

\maketitle
\begin{abstract}
Guessing Random Additive Noise Decoding (GRAND) is a universal decoding algorithm that can be used to  perform maximum likelihood decoding. 
It attempts to find the errors introduced by the channel by generating a sequence of possible error vectors in order of likelihood of occurrence and applying them to the received vector.
Ordered reliability bits GRAND (ORBGRAND) integrates soft information received from the channel to refine the error vector sequence.
In this work, ORBGRAND is modified to produce a soft output, to enable its use as an iterative soft-input soft-output (SISO) decoder.
Three techniques specific to iterative GRAND-based decoding are then proposed to improve the error-correction performance and decrease computational complexity and latency.
Using the OFEC code as a case study, the proposed techniques are evaluated, yielding substantial performance gain and astounding complexity reduction of 48\% to 85\% with respect to the baseline SISO ORBGRAND.

\end{abstract}


\IEEEpeerreviewmaketitle

\section{Introduction}\label{sec:intro}

Guessing Random Additive Noise Decoding (GRAND) \cite{GRAND_first}, whose earliest roots can be found in \cite{GRAND_marc}, has been recently proposed as a practical way to perform maximum likelihood (ML) or near-ML decoding.
It can be implemented with low complexity, and is a universal decoding algorithm, as it can decode any type of code. 
Instead of using the code structure to detect and correct errors, GRAND tries to guess the error vector applied to the transmitted codeword; ML decoding can be achieved by scheduling the estimated error vectors from the more likely to the least likely.

Some evolutions on the GRAND concept integrate soft information received from the channel to devise a refined error vector schedule \cite{SGRAND_first,ORBGRAND_first}.
Ordered Reliability Bits GRAND (ORBGRAND) \cite{ORBGRAND_first} sorts the channel soft information in order of reliability, and schedules the error patterns based on their logistic weights.
Improvements to ORBGRAND have been recently proposed \cite{iLWO,ORBGRAND_cap}, together with practical implementations of hardware decoders \cite{LUTGRAND,ORBARCH_J} that showcase different trade-offs between performance and complexity.

The intended targets of GRAND-based decoding are short, high rate codes. 
These are often used as component codes in concatenated forward error-correction (FEC) schemes like product codes \cite{elias} and staircase codes \cite{Staircase}, that are usually decoded iteratively.
The enhanced error-correction performance of soft-input soft-output (SISO) decoders can greatly reduce the number of iterations necessary to achieve the desired bit error rate (BER), and lower the error floor \cite{SDSC}.
However, no incarnation of GRAND is naturally SISO.

In this paper, ORBGRAND is modified to enable iterative SISO decoding. 
Starting from a straightforward implementation, three optimizations are then proposed, that address the iterative use of SISO ORBGRAND to enhance error-correction performance and decrease computational complexity and latency.
The OFEC code \cite{openROADM} is used as a case study to show the impact of the proposed techniques on BER and complexity, demonstrating substantial performance improvement and tremendous complexity reduction of 48\% to 85\% with respect to the baseline SISO ORBGRAND.

The rest of the paper is organized as follows. 
Section \ref{sec:prel} introduces GRAND-based decoding, together with the basic details on the OFEC.
Section \ref{sec:SISO} details the modifications necessary to make ORBGRAND a SISO algorithm, while in Section \ref{sec:SISOenhance} the proposed enhancements are described.
Finally, Section \ref{sec:conc} draws the conclusion.

\section{Preliminaries}\label{sec:prel}
\subsection{GRAND-based decoding}%
Let $\mathcal{C}$ be a binary linear block code code, with $\mathbf{G}$ its $k \times n$ generator matrix and $\mathbf{H}$ the $(n-k) \times n$ parity check matrix.
The $k$-bit vector $\mathbf{u}$ is encoded into the $n$-bit codeword $\mathbf{x}$  through $\mathbf{x}=\mathbf{u}\cdot \mathbf{G}$.
The codebook of $\mathcal{C}$ is composed of all the $2^k$ possible $\mathbf{x}$, that satisfy the following:
\begin{equation}
\forall \mathbf{x} \in \mathcal{C},\mathbf{H} \cdot \mathbf{x}^{\rm T} = \mathbf{0}~,
\end{equation}
where $\mathbf{0}$ is the all-zero vector. 
Let $\mathbf{x}$ be transmitted over a noisy channel, and let $\mathbf{y}$ be  the received soft-value vector, with the associated hard-decision vector ${\rm HD}(\mathbf{y}) = \mathbf{x} \oplus \mathbf{e}$, where $\mathbf{e}$ is the error vector applied by the channel.
In case $\mathbf{H} \cdot {\rm HD}(\mathbf{y})^{\rm T} \neq \mathbf{0}$, errors have been detected.
In the remainder of this work, the elements of $\mathbf{y}$ are logarithmic likelihood ratios (LLRs), and binary phase shift keying (BPSK) signaling is assumed, where zeros are mapped to +1 and ones are mapped to -1.

The Guessing Random Additive Noise Decoding (GRAND) algorithm proposed in \cite{GRAND_first} tries to find the error vector $\mathbf{e}$ applied by the channel.
An error vector $\mathbf{\hat{e}}$ is generated, allowing the computation of $\hat{\mathbf{x}}={\rm HD}(\mathbf{y}) \oplus \mathbf{\hat{e}}$.
The codebook is queried by computing $\mathbf{H} \cdot \hat{\mathbf{x}}^{\rm T}$: if the result is $\mathbf{0}$, the decoding is successful, otherwise a new $\mathbf{\hat{e}}$ is generated. 
GRAND repeats the above steps until a valid codeword is found; however, GRAND with abandonment (GRANDAB, \cite{GRAND_first}) terminates the decoding process after a maximum number of codebook queries $Q_{max}$. 
The $Q_{max}$ parameter has a large impact on the error-correction performance, average and worst case latency, and throughput of not only GRANDAB, but all practical versions of GRAND-based decoding.

To achieve ML decoding, the attempted $\mathbf{\hat{e}}$ should be scheduled in descending order of probability. 
For binary symmetric channels the optimal ordering is of increasing Hamming weight $HW$ of $\mathbf{\hat{e}}$, but this is not the case for additive white Gaussian noise channels (AWGN). 
Using the soft information vector $\mathbf{y}$ instead of ${\rm HD}(\mathbf{y})$ only, ORBGRAND \cite{ORBGRAND_first} infers a refined schedule for $\mathbf{\hat{e}}$.
The elements of vector $\mathbf{y}$ are sorted in ascending order of reliability, i.e. of increasing magnitude in case of LLRs, leading to a permutation $\pi$ of bit indices and the associated sorted vector $\pi(\mathbf{y})$. 
Error vectors $\mathbf{\hat{e}}$ are applied to $\pi(\mathbf{y})$ in ascending \emph{logistic weight order} (LWO). 
Given the ordered vector $\mathbf{v} = (v_0,\dots,v_{HW-1})$ that contains the indices of the nonzero entries of $\mathbf{\hat{e}}$, and its length $HW$, then the logistic weight $LW$ of $\mathbf{\hat{e}}$ can be computed as
\begin{equation}\label{eq:LW}
LW(\mathbf{\hat{e}}) = \sum_{i=0}^{HW-1} (v_i+1)~.
\end{equation}
Error vectors with the same $LW$ can be scheduled in any order, for example in ascending $HW$. 
In \cite{iLWO}, an improved logistic weight order (iLWO) schedule was proposed, where the weights of $\mathbf{\hat{e}}$ are calculated as : 
\begin{equation}
iLW(\mathbf{\hat{e}}) = \sum_{i=0}^{HW-1} (i+1)\cdot(v_i+1)~.
\end{equation}
The iLWO resulting from the above $iLW$ favors error vectors with low $HW$ and potentially high $LW$ over vectors with low $LW$ but high $HW$, and substantially improves the BER of ORBGRAND.

\subsection{OFEC}

GRAND-based decoding algorithms are universal, as they can be used to decode any linear block code. 
GRAND is particularly interesting for the decoding of short, high-rate codes, for which it can achieve ML performance with limited complexity, whereas as $n$ increases the implementation cost rises significantly, while the code rates for which good performance can be obtained decrease.
Since non-universal decoding algorithms tend to benefit from long codes, FEC schemes currently used in standardized communications often consider large $n$. 
However, shorter codes are commonly used as component codes for concatenated FEC schemes, like product codes \cite{elias} and staircase codes \cite{Staircase}. 
Concatenated FEC schemes are widespread in optical communication standards, as they enable high net coding gain and high throughput.
A recent, representative example of such FEC schemes is the so-called OFEC adopted by the 400 Gb/s Open ROADM Multi-Source Agreement \cite{openROADM}, also used for 450km distance transmissions scenarios of the G.709.3 standard \cite{ITU}, and considered among candidates for 800 Gb/s communications as well \cite{800G}.
OFEC is a staircase-like code where every bit is part of two component codewords.
The component code is a two-error-correcting Bose-Chaudhuri-Hocquenghem (BCH) \cite{BCH_forney} code extended by an overall parity bit, for a total length of $n=256$ and dimension $k=239$. 
The BCH generator polynomial is $g_\mathcal{C}=~$0x18DED defined on GF($2^8$), with a field generator polynomial $g_\mathcal{F}=~$0x171.  
The OFEC is shown to be able to yield a very steep waterfall at an input BER$~\approx 0.02$, for a net coding gain of 11.1 dB \cite{openROADM}. 
This performance is achieved through three SISO decoding iterations followed by two hard-input hard-output (HIHO) iterations. 
Chase decoding \cite{CHASE} with a large number of test patterns is used in the SISO iterations, while bounded distance decoding is used to clear residual errors in the HIHO iterations.

\section{Iterative SISO decoding with ORBGRAND}\label{sec:SISO}

GRAND is an inherently HIHO decoding algorithm, as ${\rm HD}(\mathbf{y})$ is the only input the decoder considers. 
Soft GRAND and ORBGRAND \cite{SGRAND_first,ORBGRAND_first} consider soft inputs; however, the decoding output is still the vector of bits $\hat{\mathbf{x}}$, thus being soft-input hard-output (SIHO) algorithms.
While it is possible to iterate HIHO decoders, SISO iterative decoding substantially improves error-correction performance with a limited number of iterations, and helps lowering error floors. 
GRAND has been applied as an HIHO decoder for turbo product codes in \cite{IGRAND}, but no SISO decoder based on GRAND has been proposed so far. 

While some decoding algorithms are inherently SISO, like the belief propagation algorithm often used to decode low-density parity-check codes \cite{LDPC}, HIHO and SIHO algorithms need to be modified to be used in SISO iterative decoding. 
In particular, inspired by the method initially proposed in \cite{block_turbo} to perform iterative SISO Chase decoding on turbo product codes, SIHO algorithms can be adapted to SISO \cite{PPC}.
Let $\hat{\mathbf{y}}^{i-1}$ be the soft output of iteration $i-1$, and thus the input of iteration $i$. 
If $i=0$, then $\hat{\mathbf{y}}^{i-1} = \mathbf{y}$.
After the decoding process, given the decoder first codeword candidate $\hat{\mathbf{x}}$, i.e. the original output of the SIHO algorithm, let us assume that a second, less likely competitor candidate $\hat{\mathbf{x}}^C$ is found as well.
The distance metric $\mathcal{M}$ ($\mathcal{M}_C$) expressing the Euclidean distance between $\hat{\mathbf{x}}$ ($\hat{\mathbf{x}}^C$) and the hard decision over the input ${\rm HD}(\hat{\mathbf{y}}^{i-1})$ is computed as
\begin{equation}
\mathcal{M} = 2\sum_{j=\hat{x}_j\neq {\rm HD}(\hat{y}^{i-1}_j)} |\hat{y}^{i-1}_j|~,
\label{eq:M}
\end{equation}
where $0\le j < n$, and the difference between the two is defined as 
\begin{equation}
\delta = \frac{\mathcal{M}_C-\mathcal{M}}{2}~.
\label{eq:delta}
\end{equation}
The decoder soft output $\hat{\mathbf{y}}^i$ at iteration $i$ can be computed as 
\begin{equation}
\hat{y}^i_j = y_j + \alpha^i \cdot \epsilon_j~,
\label{eq:rp}
\end{equation}
where $\alpha^i$ is a scaling factor associated to iteration $i$, and the extrinsic information vector $\mathbf{\epsilon}$ is calculated as follows:
\begin{equation}
\epsilon_j =
  \begin{cases}
    \gamma^i \cdot \delta \cdot(1-2\hat{x}_j) & \text{if } \hat{x}_j = \hat{x}^C_j~,\\
    \delta \cdot(1-2\hat{x}_j) - \hat{y}_j^{i-1} & \text{if } \hat{x}_j \neq \hat{x}^C_j~,\\
    \beta^i \cdot (1-2\hat{x}_j) & \text{if no } {\hat{\mathbf{x}}^C} \text{ was found}~.\\
  \end{cases} \label{eq:ei}
\end{equation}
In (\ref{eq:ei}), $\beta^i$ and $\gamma^i$ are iteration-dependent scaling factors, while $(1-2\hat{x}_j)$ maps bits to $\pm 1$. 
The third case takes in account the possibility that $\hat{\mathbf{x}}^C$ could not be found by the decoder, its intended meaning being that the decoder is extremely confident of the correctness of $\hat{\mathbf{x}}$.

In the remainder of the paper, the decoding of OFEC is always assumed to involve three SISO iterations followed by two HIHO iterations. 
The HIHO iterations are performed through GRANDAB, attempting all error vectors with $HW\le2$ ($Q_{max}\approx 2^{15}$).
The SISO iterations are performed through ORBGRAND, with the soft output computed according to (\ref{eq:M})-(\ref{eq:ei}), and the LLRs being quantized with 4 bits.
In case of inherently parallel decoding algorithms like Chase and list decoders, multiple codeword candidates are returned with high probability, and the competitor candidate codeword $\hat{\mathbf{x}}^C$ is readily available.
On the other hand, ORBGRAND is conceptually serial, and it stops as soon as $\hat{\mathbf{x}}$ is found. 
As a first step towards effective SISO iterative decoding, ORBGRAND is made to continue the decoding process after $\hat{\mathbf{x}}$ is found until also $\hat{\mathbf{x}}^C$ is identified, or until the maximum number of codebook queries $Q_{max}$ is reached.
In the following Section, this straightforward approach is refined to improve error-correction performance and reduce complexity and latency.

\begin{figure}[t!]
        \centering
		  \begin{tikzpicture}
  \pgfplotsset{
    label style = {font=\fontsize{9pt}{7.2}\selectfont},
    tick label style = {font=\fontsize{7pt}{7.2}\selectfont}
  }

\begin{axis}[
	scale = 1,
    ymode=log,
    xlabel={$E_b/N_0$ [\text{dB}]}, xlabel style={yshift=0.4em},
    ylabel={BER}, ylabel style={yshift=-0.75em},
    grid=both,
    ymajorgrids=true,
    xmajorgrids=true,
    grid style=dashed,
    mark options=solid,
    width=1\columnwidth,
    thick,
        xmin=6.3,
        xmax=6.8,
        ymin=1e-7,
    mark size=3,
    legend style={
      anchor={center},
      cells={anchor=west},
      mark options=solid,
      column sep= 2mm,
      font=\fontsize{7pt}{7.2}\selectfont,
    },
    legend to name=BER_iters,
    legend columns=2,
]

%

\addplot[
    color=black,
    mark=x,
    thick,
    mark size=3,
]
table {
6.0		0.0218272
6.1		0.0170545
6.2		0.0109833
6.3		0.00509627
6.4		0.00108545
6.5		3.28678e-05
6.6		1.00443e-06
6.7	    1.712e-8    
};
\addlegendentry{LWO $i=\{0,1\}$, iLWO $i=2$}
%
%
%
%

\addplot[
    color=black,
    mark=x,
    dashed,
    thick,
    mark size=3,
]
table {
6.0		0.0224945
6.1		0.0182358
6.2		0.0131796
6.3		0.00645532
6.4		0.00148055
6.5		2.27051e-05
6.6		1.89052e-07
6.65	1.2996e-08
};
\addlegendentry{LWO $i=0$, iLWO $i=\{1,2\}$}

%

\addplot[
    color=CornflowerBlue,
    mark=o,
    thick,
    mark size=3,
]
table {
6.0		0.0215293
6.05	0.0194217
6.1		0.0168311
6.15	0.0141855
6.2		0.0114859
6.25	0.00859424
6.3		0.00564941
6.35	0.00329708
6.4		0.00147673
6.45	0.000483846
6.5		8.36589e-05
6.55	1.46891e-05
6.6		3.82487e-06
6.65	7.55498e-07
6.7		1.0588e-7
};
\addlegendentry{LWO $i=\{0,1,2\}$}

\addplot[
    color=CornflowerBlue,
    mark=o,
	dashed,
    thick,
    mark size=3,
]
table {
6.0		0.0232982
6.05	0.021313
6.1		0.0192729
6.15	0.0168558
6.2		0.0142059
6.25	0.0113095
6.3		0.00797734
6.35	0.00521452
6.4		0.00205404
6.45	0.000619792
6.5		2.81576e-05
6.55	4.84212e-06
6.6		5.6276e-7
6.65	6e-8
};
\addlegendentry{iLWO $i=\{0,1,2\}$}

\end{axis}
\end{tikzpicture}
		  \\
       \ref{BER_iters}
    \caption{BER for $Q_{max}=2^{13}$ for LWO/iLWO schedules at different iterations.}
    \label{fig:BERiters}
\end{figure}
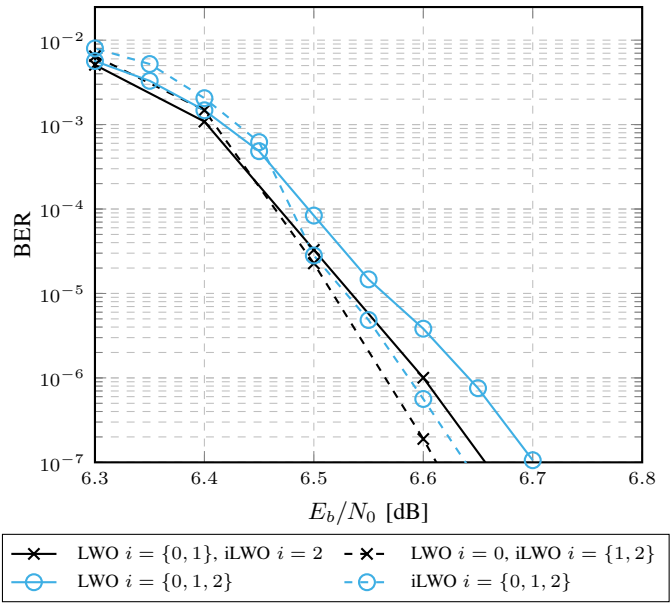

\section{Enhanced low-complexity SISO ORBGRAND}\label{sec:SISOenhance}

\subsection{Iteration-dependent error vector schedule} \label{subsec:iter}

\begin{figure}[t!]
        \centering
		  \begin{tikzpicture}
  \pgfplotsset{
    label style = {font=\fontsize{9pt}{7.2}\selectfont},
    tick label style = {font=\fontsize{7pt}{7.2}\selectfont}
  }

\begin{axis}[
	scale = 1,
    ymode=log,
    xlabel={$E_b/N_0$ [\text{dB}]}, xlabel style={yshift=0.4em},
    ylabel={BER}, ylabel style={yshift=-0.75em},
    grid=both,
    ymajorgrids=true,
    xmajorgrids=true,
    grid style=dashed,
    mark options=solid,
    width=1\columnwidth,
    thick,
        xmin=6,
        xmax=6.8,
        ymin=1e-7,
    mark size=3,
    legend style={
      anchor={center},
      cells={anchor=west},
      mark options=solid,
      column sep= 2mm,
      font=\fontsize{7pt}{7.2}\selectfont,
    },
    legend to name=BER_genie,
    legend columns=2,
]

\addplot[
    color=BurntOrange,
    mark=square,
    thick,
    mark size=3,
]
table {
6.0		0.0213861
6.05	0.0193158
6.1		0.0171342
6.15	0.0147073
6.2		0.0122152
6.25	0.00925643
6.3		0.00656999
6.35	0.00428532
6.4		0.00206567
6.45	0.000832357
6.5		0.000178304
6.55	3.19417e-05
6.6		5.8431e-06
6.65	1.70898e-06
6.7		1.89887e-7
};
\addlegendentry{LWO no $\hat{\mathbf{x}}^C$}

\addplot[
    color=BurntOrange,
    mark=square,
	dashed,
    thick,
    mark size=3,
]
table {
6.0		0.0236681
6.05	0.0218128
6.1		0.0192336
6.15	0.0169816
6.2		0.0146378
6.25	0.0113694
6.3		0.00837679
6.35	0.0059908
6.4		0.00288428
6.45	0.00109424
6.5		0.000184652
6.55	1.4445e-05
6.6		2.7832e-06
6.65	7.97313e-07
6.7		8.5e-8
};
\addlegendentry{iLWO no $\hat{\mathbf{x}}^C$}

\addplot[
    color=CornflowerBlue,
    mark=o,
    thick,
    mark size=3,
]
table {
6.0		0.0215293
6.05	0.0194217
6.1		0.0168311
6.15	0.0141855
6.2		0.0114859
6.25	0.00859424
6.3		0.00564941
6.35	0.00329708
6.4		0.00147673
6.45	0.000483846
6.5		8.36589e-05
6.55	1.46891e-05
6.6		3.82487e-06
6.65	7.55498e-07
6.7		1.0588e-7
};
\addlegendentry{LWO attempted $\hat{\mathbf{x}}^C$}

\addplot[
    color=CornflowerBlue,
    mark=o,
	dashed,
    thick,
    mark size=3,
]
table {
6.0		0.0232982
6.05	0.021313
6.1		0.0192729
6.15	0.0168558
6.2		0.0142059
6.25	0.0113095
6.3		0.00797734
6.35	0.00521452
6.4		0.00205404
6.45	0.000619792
6.5		2.81576e-05
6.55	4.84212e-06
6.6		5.6276e-7
6.65	6e-8
};
\addlegendentry{iLWO attempted $\hat{\mathbf{x}}^C$}

\addplot[
    color=black,
    mark=x,
    thick,
    mark size=3,
]
table {
6.0		0.0204486
6.05	0.0175342
6.1		0.0149929
6.15	0.0113122
6.2		0.00715609
6.25	0.0036005
6.3		0.00115259
6.35	0.000105632
6.4		1.13118e-05
6.45	4.32943e-07
};
\addlegendentry{LWO Chase-aided $\hat{\mathbf{x}}^C$}

\addplot[
    color=black,
    mark=x,
	dashed,
    thick,
    mark size=3,
]
table {
6.0		0.0227063
6.05	0.020226
6.1		0.0170179
6.15	0.014117
6.2		0.0100288
6.25	0.00557796
6.3		0.00196167
6.35	0.000135742
6.4		5.0727e-06
6.45	1.07274e-07
};
\addlegendentry{iLWO Chase-aided $\hat{\mathbf{x}}^C$}

\end{axis}
\end{tikzpicture}
		  \\
       \ref{BER_genie}
    \caption{BER for $Q_{max}=2^{13}$ for no, attempted, and Chase-aided search of $\hat{\mathbf{x}}^C$.}
    \label{fig:BERgenie}
\end{figure}
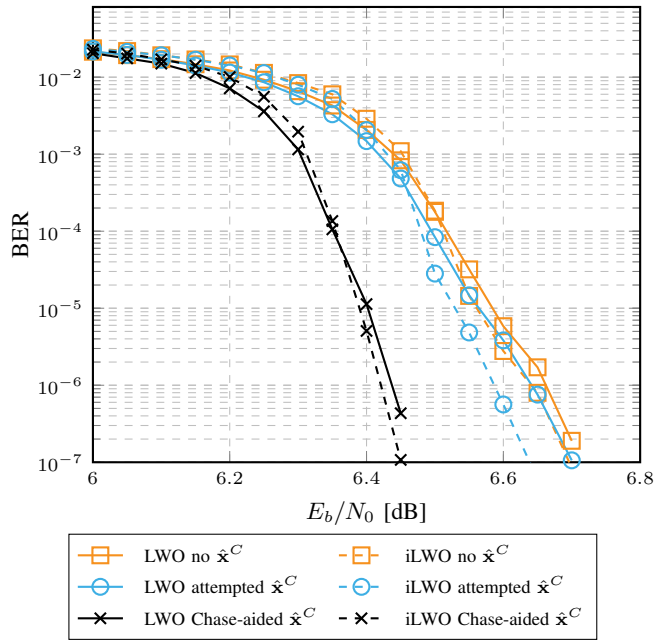

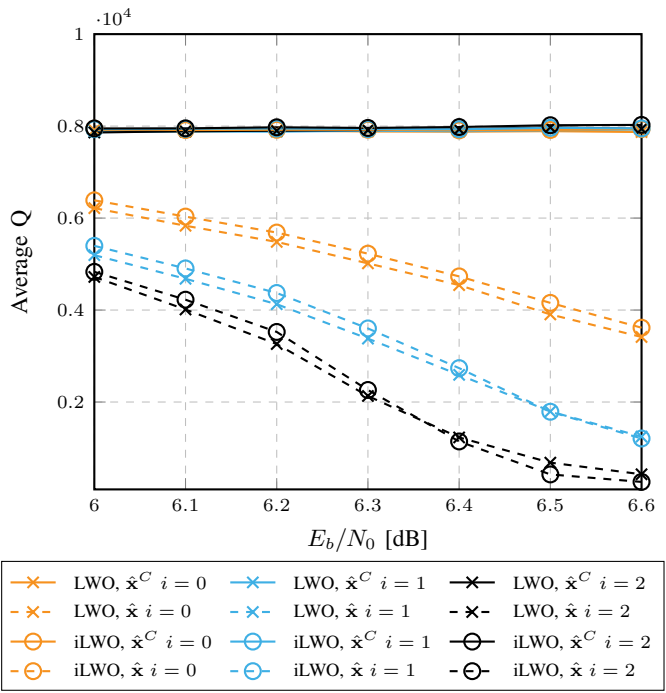
\begin{figure}[t!]
    \centering
		  \begin{tikzpicture}
  \pgfplotsset{
    label style = {font=\fontsize{9pt}{7.2}\selectfont},
    tick label style = {font=\fontsize{7pt}{7.2}\selectfont}
  }

\begin{axis}[
	scale = 1,
    xlabel={$E_b/N_0$ [\text{dB}]}, xlabel style={yshift=0.4em},
    ylabel={Average Q}, ylabel style={yshift=-0.75em},
    grid=both,
    ymajorgrids=true,
    xmajorgrids=true,
    grid style=dashed,
    mark options=solid,
    width=1\columnwidth,
    thick,
        xmin=6,
        xmax=6.6,
        ymin=100,
        ymax=10000,
    mark size=3,
    legend style={
      anchor={center},
      cells={anchor=west},
      mark options=solid,
      column sep= 1.6mm,
      font=\fontsize{7pt}{7.2}\selectfont,
    },
    legend to name=Q_DC,
    legend columns=3,
]

\addplot[
    color=BurntOrange,
    mark=x,
    thick,
    mark size=3,
]
table {
6.0 7912
6.1 7877
6.2 7906
6.3 7883
6.4 7879
6.5 7890
6.6 7871
};
\addlegendentry{LWO, $\hat{\mathbf{x}}^C$ $i=0$}

\addplot[
    color=CornflowerBlue,
    mark=x,
    thick,
    mark size=3,
]
table {
6.0 7869
6.1 7882
6.2 7879
6.3 7900
6.4 7889
6.5 7909
6.6 7926
};
\addlegendentry{LWO, $\hat{\mathbf{x}}^C$ $i=1$}

\addplot[
    color=black,
    mark=x,
    thick,
    mark size=3,
]
table {
6.0 7861
6.1 7888
6.2 7899
6.3 7914
6.4 7929
6.5 7973
6.6 7940
};
\addlegendentry{LWO, $\hat{\mathbf{x}}^C$ $i=2$}

\addplot[
    color=BurntOrange,
    mark=x,
    dashed,
    thick,
    mark size=3,
]
table {
6.0 6216
6.1 5833
6.2 5482
6.3 5015
6.4 4542
6.5 3905
6.6 3414
};
\addlegendentry{LWO, $\hat{\mathbf{x}}$ $i=0$}

\addplot[
    color=CornflowerBlue,
    mark=x,
    dashed,
    thick,
    mark size=3,
]
table {
6.0 5194
6.1 4680
6.2 4127
6.3 3379
6.4 2591
6.5 1794
6.6 1248
};
\addlegendentry{LWO, $\hat{\mathbf{x}}$ $i=1$}

\addplot[
    color=black,
    mark=x,
    dashed,
    thick,
    mark size=3,
]
table {
6.0 4720
6.1 4012
6.2 3261
6.3 2131
6.4 1230
6.5 682
6.6 429
};
\addlegendentry{LWO, $\hat{\mathbf{x}}$ $i=2$}

\addplot[
    color=BurntOrange,
    mark=o,
    thick,
    mark size=3,
]
table {
6.0 7941
6.1 7914
6.2 7920
6.3 7921
6.4 7907
6.5 7916
6.6 7935
};
\addlegendentry{iLWO, $\hat{\mathbf{x}}^C$ $i=0$}

\addplot[
    color=CornflowerBlue,
    mark=o,
    thick,
    mark size=3,
]
table {
6.0 7933
6.1 7941
6.2 7947
6.3 7936
6.4 7931
6.5 7962
6.6 7949
};
\addlegendentry{iLWO, $\hat{\mathbf{x}}^C$ $i=1$}

\addplot[
    color=black,
    mark=o,
    thick,
    mark size=3,
]
table {
6.0 7949
6.1 7951
6.2 7974
6.3 7960
6.4 7980
6.5 8019
6.6 8026
};
\addlegendentry{iLWO, $\hat{\mathbf{x}}^C$ $i=2$}

\addplot[
    color=BurntOrange,
    mark=o,
    dashed,
    thick,
    mark size=3,
]
table {
6.0 6386
6.1 6032
6.2 5685
6.3 5226
6.4 4732
6.5 4156
6.6 3616
};
\addlegendentry{iLWO, $\hat{\mathbf{x}}$ $i=0$}

\addplot[
    color=CornflowerBlue,
    mark=o,
    dashed,
    thick,
    mark size=3,
]
table {
6.0 5396
6.1 4905
6.2 4371
6.3 3600
6.4 2737
6.5 1789
6.6 1207
};
\addlegendentry{iLWO, $\hat{\mathbf{x}}$ $i=1$}

\addplot[
    color=black,
    mark=o,
    dashed,
    thick,
    mark size=3,
]
table {
6.0 4831
6.1 4225
6.2 3524
6.3 2262
6.4 1143
6.5 426
6.6 263
};
\addlegendentry{iLWO, $\hat{\mathbf{x}}$ $i=2$}

\end{axis}
\end{tikzpicture}
		  \\
         \ref{Q_DC}
\caption{Average $Q$ to obtain codeword candidate $\hat{\mathbf{x}}$ and competitor $\hat{\mathbf{x}}^C$ for $Q_{max}=2^{13}$.}
    \label{fig:Q}
\end{figure}%

The error-correction performance of ORBGRAND, given a certain $Q_{max}$, can strongly depend on the error vector schedule.
The iLWO schedule in \cite{iLWO} was proposed as an improvement to the LWO schedule of \cite{ORBGRAND_first}. 
The generally lower $HW$ of the error vectors attempted by iLWO caters to the typical input error rates when $n\approx 100$ and the output BER is low enough, i.e. an input BER$~<0.01$. 
The considered OFEC, and concatenated FEC schemes designed to be decoded iteratively in general, can however work with higher input BER (0.02 in case of OFEC). 
Moreover, the length of the OFEC component codeword is $n=256$; even though it is a concatenated FEC scheme, and it can thus be optimistically assumed that half of each decoded codeword is already error-free thanks to previous decoding, the average number of errors in the newly-received $n/2$ bits is larger than 2.5. 
The circle-marked blue curves in Figure \ref{fig:BERiters} show the BER obtained with ORBGRAND decoding, $Q_{max}=2^{13}$, for the LWO schedule (solid line) and iLWO schedule (dashed line). 
$Q_{max}=2^{13}$ is a practical limit, that strikes a good trade-off between performance, worst case latency, and throughput \cite{LUTGRAND}.
The iLWO schedule is shown to outperform LWO at BER$~\le 3\cdot 10^{-4}$, but LWO has an advantage at higher BER thanks to the averagely higher $HW$ of its error vectors.

Given that the input BER seen by iteration $i$ is approximately the output BER of iteration $i-1$, iterations are effectively working on different channels.
As such, it can be advantageous to tailor the error vector schedule to the different iterations.
The x-marked black curves in Figure \ref{fig:BERiters} have been obtained using LWO in earlier iterations and iLWO in later ones. 
It can be seen that using LWO at iteration $i=0$, followed by iLWO at $i=\{1,2\}$ outperforms all other schedule combinations.
LWO performs better than iLWO at the high input BER seen by the first iteration, while the following ones can exploit the lower input BER and further reduce it thanks to the improved performance of the iLWO schedule at low input BER.

\subsection{Efficient $\hat{\mathbf{x}}^C$ search} \label{subsec:uc}

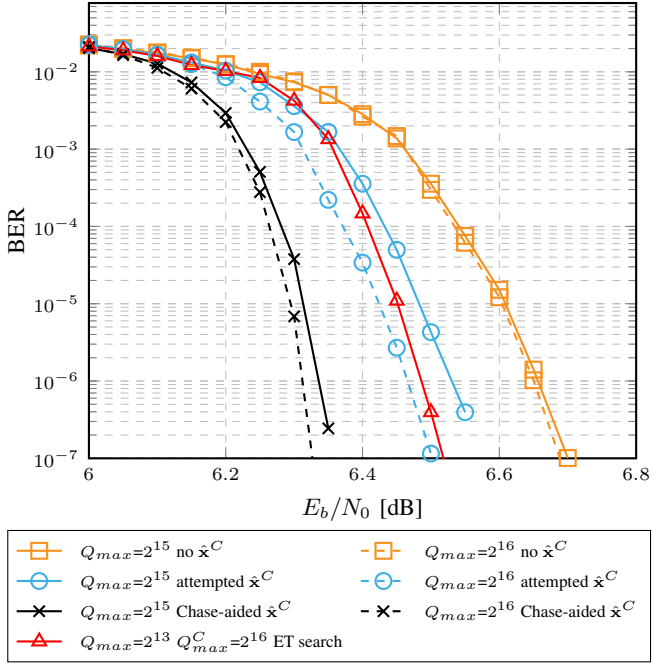
\begin{figure}[t!]
        \centering
		  \begin{tikzpicture}
  \pgfplotsset{
    label style = {font=\fontsize{9pt}{7.2}\selectfont},
    tick label style = {font=\fontsize{7pt}{7.2}\selectfont}
  }

\begin{axis}[
	scale = 1,
    ymode=log,
    xlabel={$E_b/N_0$ [\text{dB}]}, xlabel style={yshift=0.4em},
    ylabel={BER}, ylabel style={yshift=-0.75em},
    grid=both,
    ymajorgrids=true,
    xmajorgrids=true,
    grid style=dashed,
    mark options=solid,
    width=1\columnwidth,
    thick,
        xmin=6,
        xmax=6.8,
        ymin=1e-7,
    mark size=3,
    legend style={
      anchor={center},
      cells={anchor=west},
      mark options=solid,
      column sep= 1.6mm,
      font=\fontsize{6.5pt}{7.2}\selectfont,
    },
    legend to name=BER_genie_mixed,
    legend columns=2,
]

\addplot[
    color=BurntOrange,
    mark=square,
    thick,
    mark size=3,
]
table {
6.0		0.0220194
6.05	0.0196383
6.1		0.0174586
6.15	0.0149398
6.2		0.0124289
6.25	0.00937516
6.3		0.00735124
6.35	0.00502132
6.4		0.00282975
6.45	0.00137557
6.5		0.000356934
6.55	7.54395e-05
6.6		1.51367e-05
6.65	1.40381e-06
6.7		1.00887e-7
};
\addlegendentry{$Q_{max}$=$2^{15}$ no $\hat{\mathbf{x}}^C$}

\addplot[
    color=BurntOrange,
    mark=square,
	dashed,
    thick,
    mark size=3,
]
table {
6.0		0.0226191
6.05	0.0202611
6.1		0.0178733
6.15	0.015252
6.2		0.0125672
6.25	0.00999536
6.3		0.0075271
6.35	0.00502637
6.4		0.00262975
6.45	0.00147557
6.5		0.000296934
6.55	6.14395e-05
6.6		1.21367e-05
6.65	1.02281e-06
6.7		5.00887e-8
};
\addlegendentry{$Q_{max}$=$2^{16}$ no $\hat{\mathbf{x}}^C$}

\addplot[
    color=CornflowerBlue,
    mark=o,
    thick,
    mark size=3,
]
table {
6.0		0.0223206
6.05	0.0194443
6.1		0.0167295
6.15	0.0132261
6.2		0.0103643
6.25	0.00732568
6.3		0.00359424
6.35	0.0016675
6.4		0.000356608
6.45	4.9856e-5
6.5		4.28602e-06
6.55	3.94052e-07	
};
\addlegendentry{$Q_{max}$=$2^{15}$ attempted $\hat{\mathbf{x}}^C$}

\addplot[
    color=CornflowerBlue,
    mark=o,
	dashed,
    thick,
    mark size=3,
]
table {
6.0		0.0226009
6.05	0.0197176
6.1		0.0164829
6.15	0.0125968
6.2		0.00850464
6.25	0.0041233
6.3		0.00166268
6.35	0.000220622
6.4		3.40169e-05
6.45	2.69661e-06
6.5		1.15289e-07
};
\addlegendentry{$Q_{max}$=$2^{16}$ attempted $\hat{\mathbf{x}}^C$}

\addplot[
    color=black,
    mark=x,
    thick,
    mark size=3,
]
table {
6.0		0.0202638
6.05	0.0169689
6.1		0.0127853
6.15	0.00728906
6.2		0.00293774
6.25	0.000507243
6.3		3.77604e-05
6.35	2.43892e-07
};
\addlegendentry{$Q_{max}$=$2^{15}$ Chase-aided $\hat{\mathbf{x}}^C$}

\addplot[
    color=black,
    mark=x,
	dashed,
    thick,
    mark size=3,
]
table {
6.0		0.0210212
6.05	0.0164537
6.1		0.0112468
6.15	0.00607056
6.2		0.00223478
6.25	0.000277425
6.3		6.86306e-06
6.35	2.51e-9	
};
\addlegendentry{$Q_{max}$=$2^{16}$ Chase-aided $\hat{\mathbf{x}}^C$}

\addplot[
    color=red,
    mark=triangle,
    thick,
    mark size=3,
]
table {
6.0		0.0213206
6.05	0.018943
6.1		0.0160295
6.15	0.0122261
6.2		0.0102643
6.25	0.00832568
6.3		0.00423527
6.35	0.00134489
6.4		0.000147135
6.45	1.09456e-5
6.5	 	3.9527e-7
6.55	9.28e-9
};
\addlegendentry{$Q_{max}$=$2^{13}$ $Q_{max}^C$=$2^{16}$ ET search}

\end{axis}
\end{tikzpicture}
		  \\
       \ref{BER_genie_mixed}
    \caption{BER for $Q_{max}=2^{15}$ and $Q_{max}=2^{16}$ for no, attempted, and Chase-aided search of $\hat{\mathbf{x}}^C$, with LWO for $i=0$ and iLWO for $i=\{1,2\}$.}
    \label{fig:BERgenie_mixed}
\end{figure}

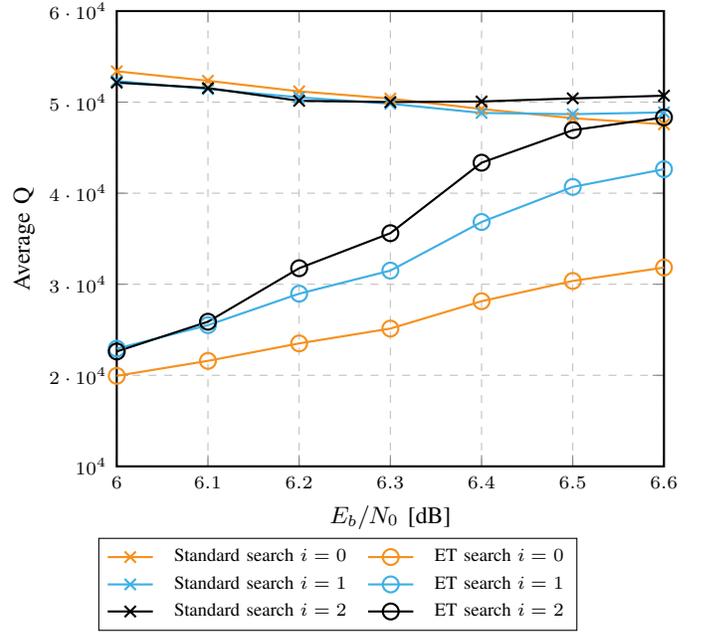
\begin{figure}[t!]
    \centering
		  \begin{tikzpicture}
  \pgfplotsset{
    label style = {font=\fontsize{9pt}{7.2}\selectfont},
    tick label style = {font=\fontsize{7pt}{7.2}\selectfont}
  }

\begin{axis}[
	scale = 1,
    xlabel={$E_b/N_0$ [\text{dB}]}, xlabel style={yshift=0.4em},
    ylabel={Average Q}, ylabel style={yshift=0em},
    grid=both,
    ymajorgrids=true,
    yminorgrids=true,
    xmajorgrids=true,
    grid style=dashed,
    mark options=solid,
    scaled y ticks = false,
    width=1\columnwidth,
    thick,
        xmin=6,
        xmax=6.6,
        ymin=10000,
        ymax=60000,
        ytick={10000,20000, 30000, 40000, 50000, 60000},
        yticklabels={$10^4$,$2\cdot10^4$,$3\cdot10^4$,$4\cdot10^4$,$5\cdot10^4$,$6\cdot10^4$},
    mark size=3,
    legend style={
      anchor={center},
      cells={anchor=west},
      mark options=solid,
      column sep= 2mm,
      font=\fontsize{7pt}{7.2}\selectfont,
    },
    legend to name=Q_best,
    legend columns=2,
]

\addplot[
    color=BurntOrange,
    mark=x,
    thick,
    mark size=3,
]
table {
6.0 53385
6.1 52345
6.2 51182
6.3 50390
6.4 49248
6.5 48236
6.6 47565
};
\addlegendentry{Standard search $i=0$}

\addplot[
    color=BurntOrange,
    mark=o,
    thick,
    mark size=3,
]
table {
6.0 19940
6.1 21585
6.2 23505
6.3 25132
6.4 28137
6.5 30364
6.6 31833
};
\addlegendentry{ET search $i=0$}

\addplot[
    color=CornflowerBlue,
    mark=x,
    thick,
    mark size=3,
]
table {
6.0 52320
6.1 51438
6.2 50525
6.3 49824
6.4 48814
6.5 48676
6.6 48875
};
\addlegendentry{Standard search $i=1$}

\addplot[
    color=CornflowerBlue,
    mark=o,
    thick,
    mark size=3,
]
table {
6.0 22926
6.1 25508
6.2 28963
6.3 31497
6.4 36835
6.5 40689
6.6 42634
};
\addlegendentry{ET search $i=1$}

\addplot[
    color=black,
    mark=x,
    thick,
    mark size=3,
]
table {
6.0 52145
6.1 51544
6.2 50146
6.3 50009
6.4 50070
6.5 50414
6.6 50705
};
\addlegendentry{Standard search $i=2$}

\addplot[
    color=black,
    mark=o,
    thick,
    mark size=3,
]
table {
6.0 22616
6.1 25883
6.2 31758
6.3 35606
6.4 43350
6.5 46920
6.6 48325
};
\addlegendentry{ET search $i=2$}

\end{axis}
\end{tikzpicture}
		  \\
         \ref{Q_best}
\caption{Total average $Q$ for ET search with $Q_{max}=\{2^{13},2^{16}\}$ and for standard search with $Q_{max}=2^{16}$, with LWO for $i=0$ and iLWO for $i=\{1,2\}$.}
    \label{fig:Q_best}
\end{figure}%

As stated earlier in this Section, ORBGRAND does not naturally provide multiple codeword candidates, being focused on the identification of $\hat{\mathbf{x}}$. 
It could be tempting to simply ignore $\hat{\mathbf{x}}^C$, and always compute $\mathbf{\epsilon}$ according to the third case in (\ref{eq:ei}).
However, this choice results in strong BER degradation with respect to the optimal case, since the reliability of $\hat{\mathbf{x}}$ is not computed relatively to one of its closest competitors, and is thus often overestimated or underestimated.
Figure \ref{fig:BERgenie} plots the BER for three different decoding choices when the same error vector schedule is applied to all soft iterations, with $Q_{max}=2^{13}$.
The square-marked orange curves show the BER when $\hat{\mathbf{x}}^C$ is never identified, while in the x-marked black curves $\hat{\mathbf{x}}^C$ is instead found through Chase decoding, with a success rate of more than 95\%. 
It can be seen that the availability of $\hat{\mathbf{x}}^C$ allows for a substantial gain in performance.
Unfortunately, simply continuing the decoding after $\hat{\mathbf{x}}$ has been found leads to very low success rates of finding $\hat{\mathbf{x}}^C$ within $Q_{max}$ ($<10\%$ at all iterations). 
Such low availability of $\hat{\mathbf{x}}^C$ is reflected on the circle-marked blue curves in Figure \ref{fig:BERgenie}, that show only marginal improvement with respect to the orange curves. 
Consequently, as portrayed by the solid curves in Figure \ref{fig:Q}, the average number of queries $Q$ performed in the search of $\hat{\mathbf{x}}^C$ is very close to $Q_{max}$, even when the average $Q$ to find $\hat{\mathbf{x}}$ decreases as $i$ and $E_b/N_0$ increase.
Chase decoding can be mimicked by an ad-hoc ORBGRAND-GRAND mixed schedule, that however is code-specific (as it requires knowledge of code characteristics like the innate error correction capability of the component code) and requires a very large number of queries, $Q_{max}\approx 2^{20}$ in case of OFEC.
Thus, achieving a high success rate of finding $\hat{\mathbf{x}}^C$ in this way is only feasible in presence of very relaxed latency constraints, and at the cost of the universality of the GRAND-based decoder.

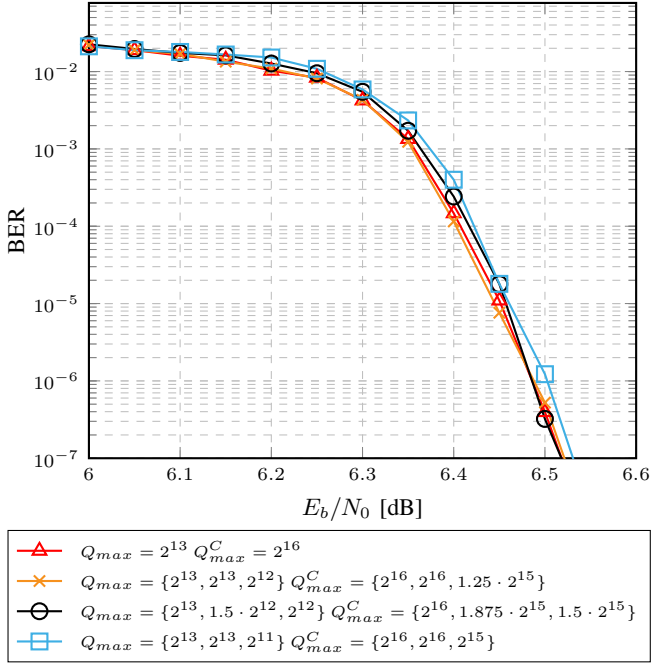
\begin{figure}[t!]
        \centering
		  \begin{tikzpicture}
  \pgfplotsset{
    label style = {font=\fontsize{9pt}{7.2}\selectfont},
    tick label style = {font=\fontsize{7pt}{7.2}\selectfont}
  }

\begin{axis}[
	scale = 1,
    ymode=log,
    xlabel={$E_b/N_0$ [\text{dB}]}, xlabel style={yshift=0.4em},
    ylabel={BER}, ylabel style={yshift=-0.75em},
    grid=both,
    ymajorgrids=true,
    xmajorgrids=true,
    grid style=dashed,
    mark options=solid,
    width=1\columnwidth,
    thick,
        xmin=6.0,
        xmax=6.6,
        ymin=1e-7,
    mark size=3,
    legend style={
      anchor={center},
      cells={anchor=west},
      mark options=solid,
      column sep= 1.6mm,
      font=\fontsize{6.5pt}{7.2}\selectfont,
    },
    legend to name=BER_diffmax,
    legend columns=1,
]

\addplot[
    color=red,
    mark=triangle,
    thick,
    mark size=3,
]
table {
6.0		0.0213206
6.05	0.018943
6.1		0.0160295
6.15	0.0142261
6.2		0.0102643
6.25	0.00832568
6.3		0.00423527
6.35	0.00134489
6.4		0.000147135
6.45	1.09456e-5
6.5	 	3.9527e-7
6.55	9.28e-9
};
\addlegendentry{$Q_{max}=2^{13}$ $Q_{max}^C=2^{16}$}

\addplot[
    color=BurntOrange,
    mark=x,
    thick,
    mark size=3,
]
table {
6.0		0.02203206
6.05	0.01888943
6.1		0.0169295
6.15	0.013422261
6.2		0.01092643
6.25	0.00812568
6.3		0.00430876
6.35	0.0012334
6.4		0.000115316
6.45	7.60905e-06
6.5		5.23158e-07
6.55	1.08e-8
};
\addlegendentry{$Q_{max}=\{2^{13},2^{13},2^{12}\}$ $Q_{max}^C=\{2^{16},2^{16},1.25\cdot2^{15}\}$}

\addplot[
    color=black,
    mark=o,
    thick,
    mark size=3,
]
table {
6.0		0.022503206
6.05	0.019588943
6.1		0.0174295
6.15	0.01622261
6.2		0.012702643
6.25	0.009512568
6.3		0.00547266
6.35	0.00171647
6.4		0.000242025
6.45	1.78658e-05
6.5		3.2213e-07
6.55	1.27e-8
};
\addlegendentry{$Q_{max}=\{2^{13},1.5\cdot2^{12},2^{12}\}$ $Q_{max}^C=\{2^{16},1.875\cdot2^{15},1.5\cdot2^{15}\}$}

\addplot[
    color=CornflowerBlue,
    mark=square,
    thick,
    mark size=3,
]
table {
6.0		0.0212003206
6.05	0.0188943
6.1		0.0179295
6.15	0.01662261
6.2		0.015202643
6.25	0.0108612568
6.3		0.00587484
6.35	0.00233398
6.4		0.000399447
6.45	1.7985e-05
6.5		1.2207e-06
6.55	2.10e-8
};
\addlegendentry{$Q_{max}=\{2^{13},2^{13},2^{11}\}$ $Q_{max}^C=\{2^{16},2^{16},2^{15}\}$}

\end{axis}
\end{tikzpicture}
		  \\
       \ref{BER_diffmax}
    \caption{BER for iteration-dependent $Q_{max}$ and $Q_{max}^C$ ET search, with LWO for $i=0$ and iLWO for $i=\{1,2\}$.}
    \label{fig:BER_diffmax}
\end{figure}

Figure \ref{fig:BERgenie_mixed} plots the BER for $Q_{max}=2^{15}$ and $Q_{max}=2^{16}$ for no, attempted, and Chase-aided search of $\hat{\mathbf{x}}^C$, with LWO for $i=0$ and iLWO for $i=\{1,2\}$. 
The square-marked orange curves do not show improvement with respect to the same curves with $Q_{max}=2^{13}$ (see Figure \ref{fig:BERgenie}). 
This is an important observation, because it reveals that if $\hat{\mathbf{x}}^C$ is ignored, a higher probability of finding $\hat{\mathbf{x}}$ does not help improving the BER.
On the other hand, the potential gain in presence of $\hat{\mathbf{x}}^C$ is larger, as indicated by the x-marked black curves.
As $Q_{max}$ increases and the decoding continues after the identification of $\hat{\mathbf{x}}$, the probability of finding $\hat{\mathbf{x}}^C$ rises to more than 25\% for $Q_{max}=2^{15}$ and to more than 45\% for $Q_{max}=2^{16}$.
This is due to the fact that at low enough BER, the average $Q$ required to find $\hat{\mathbf{x}}$ does not increase significantly with $Q_{max}$; consequently, the increasingly numerous remaining $Q_{max}-Q$ queries can be used to search for $\hat{\mathbf{x}}^C$.
The circle-marked blue curves in Figure \ref{fig:BERgenie_mixed} reflect this enhanced success rate, being closer to the Chase-aided curves than their counterpart in Figure \ref{fig:BERgenie}.
The average $Q$ required to identify $\hat{\mathbf{x}}^C$ is nevertheless very high, being close to 27000 for $Q_{max}=2^{15}$ and to 50000 for $Q_{max}=2^{16}$, much higher than that of $\hat{\mathbf{x}}$.
Thus, if the search for $\hat{\mathbf{x}}$ consumes most of the available $Q_{max}$ queries, it is very unlikely that $\hat{\mathbf{x}}^C$ will be found. 
Moreover, as can be observed from the orange curves in Figure \ref{fig:BERgenie} and \ref{fig:BERgenie_mixed}, an $\hat{\mathbf{x}}$ identified after a large number of queries does not contribute to BER reduction if not paired with $\hat{\mathbf{x}}^C$. 
Following the above observations, the search for $\hat{\mathbf{x}}^C$ can be made more efficient.
A different maximum number of queries is allocated to the search for the two codeword candidates, respectively $Q_{max}$ and $Q_{max}^C$. 
If $\hat{\mathbf{x}}$ is found within $Q_{max}$, then $\hat{\mathbf{x}}^C$ can be searched until up to $Q_{max}^C$ total queries have been attempted.
On the other hand, if $\hat{\mathbf{x}}$ is not found within $Q_{max}$ queries, the decoding stops. 
This method is a kind of early termination (ET), and it is thus labeled \emph{ET search}.
The triangle-marked red curve in Figure \ref{fig:BERgenie_mixed} has been obtained with the aforementioned ET search with $Q_{max}=2^{13}$ and $Q_{max}^C=2^{16}$; it can be seen that its performance tends to that of the standard search approach with $Q_{max}=2^{16}$.
However, the average $Q$ of the ET search is substantially lower, as it can be observed in Figure \ref{fig:Q_best}. 
Early iterations in particular can greatly benefit from ET even at high $E_b/N_0$.
At $E_b/N_0=6.5$dB, where both the standard and the proposed ET search have BER$~<10^{-6}$, the ET search yields between 7\% ($i=2$) to 37\% ($i=0$) lower average $Q$.

\subsection{Iteration-dependent $Q_{max}$ and $Q_{max}^C$} \label{subsec:iterQ}

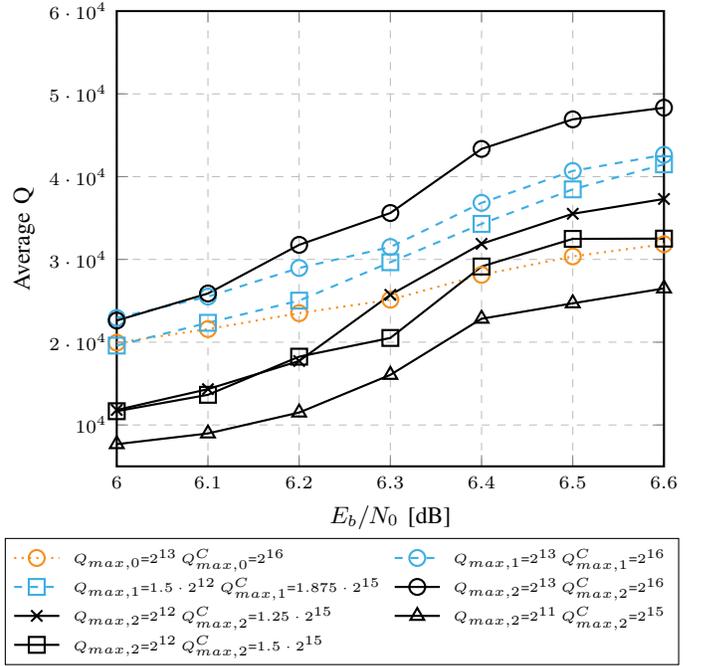
\begin{figure}[t!]
    \centering
		  \begin{tikzpicture}
  \pgfplotsset{
    label style = {font=\fontsize{9pt}{7.2}\selectfont},
    tick label style = {font=\fontsize{7pt}{7.2}\selectfont}
  }

\begin{axis}[
	scale = 1,
    xlabel={$E_b/N_0$ [\text{dB}]}, xlabel style={yshift=0.4em},
    ylabel={Average Q}, ylabel style={yshift=0em},
    grid=both,
    ymajorgrids=true,
    yminorgrids=true,
    xmajorgrids=true,
    grid style=dashed,
    mark options=solid,
    scaled y ticks = false,
    width=1\columnwidth,
    thick,
        xmin=6,
        xmax=6.6,
        ymin=5000,
        ymax=60000,
        ytick={10000,20000, 30000, 40000, 50000, 60000},
        yticklabels={$10^4$,$2\cdot10^4$,$3\cdot10^4$,$4\cdot10^4$,$5\cdot10^4$,$6\cdot10^4$},
    mark size=3,
    legend style={
      anchor={center},
      cells={anchor=west},
      mark options=solid,
      column sep= 1mm,
      font=\fontsize{5.5pt}{7.2}\selectfont,
    },
    legend to name=Q_diffmax,
    legend columns=2,
]

\addplot[
    color=BurntOrange,
    mark=o,
    dotted,
    thick,
    mark size=3,
]
table {
6.0 19940
6.1 21585
6.2 23505
6.3 25132
6.4 28137
6.5 30364
6.6 31833
};
\addlegendentry{$Q_{max,0}$=$2^{13}$ $Q_{max,0}^C$=$2^{16}$}

\addplot[
    color=CornflowerBlue,
    mark=o,
    dashed,
    thick,
    mark size=3,
]
table {
6.0 22926
6.1 25508
6.2 28963
6.3 31497
6.4 36835
6.5 40689
6.6 42634
};
\addlegendentry{$Q_{max,1}$=$2^{13}$ $Q_{max,1}^C$=$2^{16}$}

\addplot[
    color=CornflowerBlue,
    mark=square,
    dashed,
    thick,
    mark size=3,
]
table {
6.0 19592
6.1 22344
6.2 25021
6.3 29660
6.4 34290
6.5 38464
6.6 41488
};
\addlegendentry{$Q_{max,1}$=$1.5\cdot2^{12}$ $Q_{max,1}^C$=$1.875\cdot2^{15}$}

\addplot[
    color=black,
    mark=o,
    thick,
    mark size=3,
]
table {
6.0 22616
6.1 25883
6.2 31758
6.3 35606
6.4 43350
6.5 46920
6.6 48325
};
\addlegendentry{$Q_{max,2}$=$2^{13}$ $Q_{max,2}^C$=$2^{16}$}

\addplot[
    color=black,
    mark=x,
    thick,
    mark size=3,
]
table {
6.0 11811
6.1 14331
6.2 17731
6.3 25719
6.4 31885
6.5 35513
6.6 37301
};
\addlegendentry{$Q_{max,2}$=$2^{12}$ $Q_{max,2}^C$=$1.25\cdot2^{15}$}

\addplot[
    color=black,
    mark=triangle,
    thick,
    mark size=3,
]
table {
6.0 7681
6.1 8974
6.2 11500
6.3 16054
6.4 22835
6.5 24698
6.6 26494
};
\addlegendentry{$Q_{max,2}$=$2^{11}$ $Q_{max,2}^C$=$2^{15}$}

\addplot[
    color=black,
    mark=square,
    thick,
    mark size=3,
]
table {
6.0 11662
6.1 13644
6.2 18252
6.3 20512
6.4 29142
6.5 32481
6.6 32511
};
\addlegendentry{$Q_{max,2}$=$2^{12}$ $Q_{max,2}^C$=$1.5\cdot2^{15}$}

\end{axis}
\end{tikzpicture}
		  \\
         \ref{Q_diffmax}
\caption{Total average $Q$ for iteration-dependent $Q_{max}$ and $Q_{max}^C$ ET search, with LWO for $i=0$ and iLWO for $i=\{1,2\}$. Dotted curves are $i=0$, dashed curves are $i=1$, solid curves are $i=2$.}
    \label{fig:Q_diffmax}
\end{figure}%

As mentioned in Section \ref{subsec:iter}, each iteration is effectively working on a different channel, and thus benefits from an ad-hoc error vector schedule. 
In the same way, the maximum number of queries $Q_{max}$ ($Q_{max}^C$) should be adapted to the different conditions each iteration works with.
In particular, given that as $i$ increases the channel improves, $Q_{max}$ and $Q_{max}^C$ can be reduced at later iterations without substantial BER loss.
Figure \ref{fig:BER_diffmax} plots the BER for ET search, where $Q_{max,i}\ge Q_{max,i+1}$ and $Q_{max,i}^C\ge Q_{max,i+1}^C$.
In Figure \ref{fig:BER_diffmax}, the triangle-marked red curve has been obtained with $Q_{max}=2^{13}$ and $Q_{max}^C=2^{16}$ at all iterations, and serves as a reference.
With $Q_{max,2}=2^{12}$ and $Q_{max,2}^C=1.25\cdot 2^{15}$ (x-marked orange curve), there is no discernible disadvantage with respect to the reference.
If the maximum number of queries is reduced at $i=1$ as well (circle-marked black curve), a slightly more substantial loss can be observed, that nevertheless converges to the reference curve at lower BER.
With $Q_{max,2}=2^{11}$ and $Q_{max,2}^C=2^{15}$ (square-marked blue curve), a 0.01dB loss is observed at BER$=10^{-7}$.
Thanks to this approach, the average $Q$ can be substantially reduced, as it can be observed in Figure \ref{fig:Q_diffmax}. 
Of particular interest are the solid black curves, that refer to $i=2$ for the same cases as Figure \ref{fig:BER_diffmax}.
They are shown to yield 23-46\% average $Q$ reduction at high SNR and 50-66\% at low SNR with respect to the already efficient ET search is observed. 
Compared to the standard search with $Q_{max}=Q_{max}^C=2^{16}$, the reduction reaches 48\% at high SNR and 85\% at low SNR.

\section{Conclusion} \label{sec:conc}

In this work, the Ordered Reliability Bits GRAND algorithm has been adapted for iterative soft-input soft-output decoding, and its base version has been enhanced for better performance and lower complexity.
Taking the OFEC code as an example, it has been shown that by adapting the error vector schedule to the different iterations, the error-correction performance of the decoder can be improved. 
An efficient search method for the second codeword candidate required for soft output computation has been proposed, that can greatly reduce the computational complexity and decoding latency at a very small performance cost. 
Finally, this method has been further simplified by adapting its parameters to the iterative nature of the decoder, leading to an overall average latency reduction that reaches values ranging between 48\% and 85\% with respect to the baseline SISO ORBGRAND.


\end{document}